\documentclass[aps,prd,amssymb,showpacs,floatfix,preprint,nofootinbib]{revtex4-1}
\usepackage{amsmath,amsfonts,graphics,epsfig,amssymb,color}

\begin{document}
\title{Polar decomposition of the Wiener measure: Schwarzian theory versus conformal quantum mechanics }

\author{Vladimir V. Belokurov}

\email{vvbelokurov@yandex.ru}

\affiliation{Lomonosov Moscow State University, Leninskie gory 1, Moscow, 119991, Russia and Institute for Nuclear
Research of the Russian Academy of Sciences, 60th October Anniversary
Prospect 7a, Moscow, 117312, Russia}

\author{Evgeniy T. Shavgulidze}

\email{shavgulidze@bk.ru}

\affiliation{Lomonosov Moscow State University, Leninskie gory 1, Moscow, 119991, Russia}



\begin{center}
\begin{abstract}
We derive the explicit form of the polar decomposition of the Wiener measure,
and obtain the equation connecting functional integrals in conformal quantum mechanics to those in the Schwarzian theory.
Using this connection we evaluate some nontrivial functional integrals in the Schwarzian theory
 and also find the fundamental solution of the Schroedinger equation in imaginary time in the model of conformal quantum mechanics.
\end{abstract}
\end{center}
\maketitle

\vspace{1cm}
\section{ Introduction}

 In recent years, the Schwarzian theory has become extremely popular.
It is given by the action
\begin{equation}
   \label{Act2}
   I=-\frac{1}{\sigma^{2}}\int \limits _{S^{1}}\,\left[ \mathcal{S}_{\varphi}(t)+2\pi^{2}\left(\varphi'(t)\right)^{2}\right]dt\,,
\end{equation}
where
\begin{equation}
   \label{Der}
\mathcal{S}_{\varphi}(t)=
\left(\frac{\varphi''(t)}{\varphi'(t)}\right)'
-\frac{1}{2}\left(\frac{\varphi''(t)}{\varphi'(t)}\right)^2
\end{equation}
is the Schwarzian derivative, and $\varphi(t)$ is an orientation preserving $(\varphi'(t)>0)$ diffeomorphism of the unit circle $
(\varphi\in Diff^{1}_{+}(S^{1}))\,.$

The Schwarzian action appears to be the effective action  in the quantum mechanical model of Majorana fermions with a random interaction
(Sachdev-Ye-Kitaev model), in the holographic description of the Jackiw-Teitelboim dilaton gravity, in open string theory and in some other models \cite{(Kit)} - \cite{(WBCK)}.

An extraordinary universality of the Schwarzian action is a consequence of its $SL(2,\textbf{R}) $ invariance. At the same time, the Schwarzian theory is not the only $SL(2,\textbf{R}) $ invariant action. This symmetry manifests itself in conformal quantum mechanics \cite{(Alfaro)}, Liouville quantum
mechanics \cite{(PP)}-\cite{(T)} and some other models that are used to describe the near-horizon geometry of Reissner-Nordstr\"{o}m black hole and $AdS_{2}/CFT_{1}$ duality, gravity near space-like singularity (see, e.g., \cite{(Kallosh)}-\cite{(Pioline)}), and other physical problems where a universal regime is reached. Therefore, the attempts \cite{(MSY)}, \cite{(BAK)} - \cite{(SW)} to connect in some way the Schwarzian theory to other conformal invariant quantum mechanical models are quite natural.

In this case, a temptation to substitute the simpler conformal action for the Schwarzian action (\ref{Act2}) may emerge. However, as we will see below, the relationship between these theories is more difficult.

Here, to connect theories with each other we propose to study the corresponding functional integration measures.

In \cite{(Shavgulidze1978)}-\cite{(Shavgulidze2000)},  the measures
\begin{equation}
   \label{MeasureCircle}
   \mu_{\sigma}(d\varphi)=\exp\left\{\frac{1}{\sigma^{2}}\int \limits _{0}^{1}\, \mathcal{S}_{\varphi}(t)\,dt  \right\}  d\varphi
   \end{equation}
on the group $ Diff^{1}_{+}(S^{1})$,
and
\begin{equation}
   \label{MeasureInterval}
   \mu_{\sigma}(d\varphi)=\frac{1}{\sqrt{\varphi'(0)\varphi'(1)}}\exp\left\{\frac{1}{\sigma^{2}}\int \limits _{0}^{1}\,\mathcal{S}_{\varphi}(\tau)d\tau +\frac{1}{\sigma^{2}}\left[\frac{\varphi''(0)}{\varphi'(0)}-\frac{\varphi''(1)}{\varphi'(1)} \right] \right\}
  d\varphi
   \end{equation}
on the group $ Diff^{1}_{+}([0,\,1])$  were constructed.

The measure (\ref{MeasureInterval}) can be generated from the Wiener measure under some special substitution of variables \cite{(Shavgulidze1978)}-\cite{(Shavgulidze2000)}.
Namely, let us consider the function $\xi\in C_{0}([0,\,1])\,$ (that is, $\xi(t)$ is a continuous function on the interval satisfying the boundary condition
$\xi(0)=0\,$).
Then, under the substitution
\begin{equation}
   \label{subst}
 \varphi(t)=\frac{\int \limits _{0}^{t}\,e^{\xi(\tau)}d\tau}{\int \limits _{0}^{1}\,e^{\xi(\eta)}d\eta }  \,,\ \ \ \ \ \xi(t)=\log\varphi'(t)-\log\varphi'(0)\,,
\end{equation}
the measure $\mu_{\sigma}(d\varphi)$ on the group $ Diff^{1}_{+}([0,\,1])$ turns into the Wiener measure  $w_{\sigma}(d\xi)$ on $C_{0}([0,\, 1])\,.$

The measures (\ref{MeasureCircle}) and (\ref{MeasureInterval}) are quasi-invariant under the action of the subgroup consisting of three times differentiable transformations $ Diff^{3}_{+}(S^{1})$ $(\, Diff^{3}_{+}([0,\,1])\,)\,.$

  Quasi-invariance means that under the action of the subgroup $G= Diff^{3}_{+}(S^{1})$  ($\,G=Diff^{3}_{+}([0,\,1])\,)$, the measure $\mu_{\sigma}(d\varphi)$ transforms to itself multiplied by a function $  \mathcal{R}_{g}(h)$ parametrized by the elements of the subgroup $g\in G \,:$
$\mu \left(\,d(g\circ h)\,\right)=\mathcal{R}^{\mu}_{g}(h)\,\mu (\,dh\,)\,.
$)

Using the quasi-invariance of the measure $\mu_{\sigma}(d\varphi)$,   we have evaluated functional integrals for the partition function and the correlation functions in the Schwarzian theory explicitly, and derive the general rules of functional integration in the theories of the Schwarzian type \cite{(BShExact)}-\cite{(BShRules)}.

  The Wiener measure on the infinite-dimensional space of continuous positive functions $C_{+}([0,\,1])$ is also known to be
  quasi-invariant under the action of $ Diff^{3}_{+}([0,\,1])$ \cite{(Shepp)}, \cite{(Shavgulidze2000)} (see also \cite{(BSh)}).

In this paper (sections \ref {sec:str}, \ref {sec:P}), we find that the two measures are connected with each other by the equation
$$
w_{\sigma}(dx)=\exp\left\{-\frac{\sigma^{2}}{8\rho^{2}}\right\}\ \left(\varphi'(0)\varphi'(1) \right)^{\frac{3}{4}}\ \mu_{\frac{2\sigma}{\rho}}(d\varphi)\ d\rho\,,
$$
$$
x\in C_{+}([0,\,1])
\,,
\ \ \ \ \ \varphi\in Diff^{1}_{+}([0,\,1])\,,
\ \ \ \ \ 0<\rho<+\infty\,,
$$
which we name "Polar decomposition of the Wiener measure".

The corresponding polar decomposition of the measure on the circle has the form
$$
w_{\sigma}(dx)=\exp\left\{-\frac{\sigma^{2}}{8\rho^{2}}\right\}\  \mu_{\frac{2\sigma}{\rho}}(d\varphi)\ d\rho\,,
$$
$$
x\in C_{+}(S^{1})
\,,
\ \ \ \ \ \varphi\in Diff^{1}_{+}(S^{1})\,,
\ \ \ \ \ 0<\rho<+\infty\,.
$$

In section \ref{sec:conSchw}, for
 $SL(2,\textbf{R}) $ invariant functionals $\Phi$ we obtain the equation
$$
\int \limits _{  C_{+}(S^{1})/SL(2,\textbf{R}) }\Phi(x)\,
\exp\left\{-\frac{1}{2\sigma^{2}}\int \limits _{S^{1}}\left[\left(x'(t) \right)^{2} -2\pi^{2}x^{2}(t) +
\frac{2g}{x^{2}(t)}\right] dt\right\}\,dx
$$
$$
=\int\limits_{0}^{+\infty}\,d\rho\,\exp\left\{-\left(\frac{\sigma^{2}}{8}+\frac{g}{\sigma^{2}}\right)\frac{1}{\rho^{2}} \right\}
$$
$$
\times\int \limits _{  Diff_{+}^{1}(S^{1})/SL(2,\textbf{R}) }\Phi(x(\rho,\,\varphi))\,
\exp\left\{\frac{\rho^{2}}{4\sigma^{2}}\int \limits _{S^{1}}\left[\mathcal{S}_{\varphi}(t)+2\pi^{2}\left(\varphi'(t)\right)^{2}
\right] dt\right\}\,d\varphi\,.
$$
It connects functional integrals in conformal quantum mechanics to corresponding functional integrals in the Schwarzian theory, and makes it possible
to improve the technique of functional integration significantly to evaluate nontrivial functional integrals in both theories.

In section \ref{sec:Anex}, we demonstrate an example of the application of above equation in conformal quantum mechanics. Using already evaluated functional integrals in the Schwarzian theory, we find the fundamental solution of the Schr\"{o}dinger equation in imaginary time in the quantum mechanical model with Calogero-type potential.

In section \ref{sec:concl}, we give the concluding remarks.

\section{Stratification of the space  $C_{+}([0,\,1])$ and polar decomposition of the Wiener measure }
\label{sec:str}

Consider the Wiener measure with the variance $\sigma$ on the space of continuous positive functions $x(t) $  on the interval $[0,\,1]\,,$ with arbitrary values $x(0) $ and $x(1) $ independent of each other. Formally it is written as
\begin{equation}
   \label{Wiener}
 w_{\sigma}(dx)=\exp\left\{ -\frac{1}{2\sigma^{2}}\int \limits _{0}^{1}\,\left(x'(t) \right)^{2}dt \right\}\ dx\,.
\end{equation}

The measure (\ref{Wiener}) is quasi-invariant under the following action of the group of diffeomorphisms $Diff^{3}_{+} ([0,\,1])$ on $C_{+}([0,\,1])$ \cite{(Shepp)}, \cite{(Shavgulidze2000)}:
\begin{equation}
   \label{fx}
x\mapsto fx\,,\ \ (fx)(t)=x\left(f^{-1}(t) \right)\frac{1}{\sqrt{\left(f^{-1}(t) \right)'}}\,,
\end{equation}
$$
x\in C_{+}([0,\,1])\,,\ \ \  f\in Diff^{3}_{+}([0,\,1])\,.
$$

There is the invariant under the action (\ref{fx}) of the group $Diff^{3}_{+} ([0,\,1])$.
 It is given by the integral
\begin{equation}
   \label{inv}
\frac{1}{\rho^{2}}=\int \limits _{0}^{1}\,\frac{1}{x^{2}(t)}dt\,.
\end{equation}

Define $\varphi\in Diff^{1}_{+} ([0,\,1])$ by the equation
\begin{equation}
   \label{fi}
\varphi^{-1}(t)=\rho^{2}\,\int \limits _{0}^{t}\,\frac{1}{x^{2}(\tau)}d\tau \,.
\end{equation}
Then $x(t)$ is expressed in terms of $\rho $ and $\varphi(t) $
\begin{equation}
   \label{xrofi}
x(t)=\rho\,\frac{1}{\sqrt{\left(\varphi^{-1}(t)\right)'}}\,.
\end{equation}
In this case, we have
\begin{equation}
   \label{intxsq}
 \int \limits _{0}^{1}\,x^{2}(t) dt =\rho^{2}\,\int \limits _{0}^{1}\,\left(\varphi'(\tau)\right)^{2}d\tau\,.
\end{equation}

Therefore, there is a one-to-one correspondence $
(\rho\,,\ \varphi)\leftrightarrow x\,,$ and the space  $C_{+}([0,\,1])$ is stratified into the orbits with different values of the invariant $\rho\,.$

For smooth $ x(t)$ $(x\in C^{1}_{+}([0,\,1]))$ and three time differentiable $ \varphi(t)$ $(\varphi\in Diff^{3}_{+}([0,\,1]))\,,$
we also obtain
\begin{equation}
   \label{intxprim}
 -\frac{1}{2\sigma^{2}}\int \limits _{0}^{1}\,\left(x'(t) \right)^{2}dt =\frac{\rho^{2}}{4\sigma^{2}}\,\left\{\int \limits _{0}^{1}\,\mathcal{S}_{\varphi}(\tau)d\tau +\left[\frac{\varphi''(0)}{\varphi'(0)}-\frac{\varphi''(1)}{\varphi'(1)} \right] \right\}\,.
\end{equation}

Thus, for the Wiener measure on the space $C_{+}([0,\,1])$ the following polar decomposition is valid:
\begin{equation}
   \label{Polar}
w_{\sigma}(dx)=\mathcal{P}_{\sigma}(\rho)\ \left(\varphi'(0)\varphi'(1) \right)^{\frac{3}{4}}\ \mu_{\frac{2\sigma}{\rho}}(d\varphi)\, d\rho\,.
\end{equation}
The normalizing factor $\mathcal{P}(\rho)$ (to be  evaluated below) determines the relative weight of the input to the measure from the path $x(t)$ with the definite value $\frac{1}{\rho^{2}}$ of the invariant (\ref{inv}).

\section{Normalizing factor in the polar decomposition of the Wiener measure}
\label{sec:P}

To find the explicit form of the density $\mathcal{P}(\rho)$, we multiply (\ref{Polar}) by the two $\delta - $functions and consider the following equality of functional integrals:
$$
\int \limits _{ C_{+}([0,\,1]) }\,\delta \left(\tilde{\rho}-\left(\int \limits _{0}^{1}\,\frac{1}{x^{2}(t)}dt \right)^{-\frac{1}{2}}\right)
\,\delta\left(x(0)-x(1)\right)\,w_{\sigma}(dx)=\int \limits _{0}^{+\infty}\,\delta (\tilde{\rho} - \rho)\,\mathcal{P}_{\sigma}(\rho)
$$
\begin{equation}
   \label{PolarFI}
\times \int \limits _{  Diff_{+}^{1}([0,\,1]) }\,\delta \left(\frac{\rho}{\sqrt{\left(\varphi^{-1}(0) \right)'}}-\frac{\rho}{\sqrt{\left(\varphi^{-1}(1) \right)'}}\right)\ \left(\varphi'(0)\varphi'(1) \right)^{\frac{3}{4}}\ \mu_{\frac{2\sigma}{\rho}}(d\varphi)\,d\rho
\end{equation}

First, we evaluate the functional integral in the right-hand side of (\ref{PolarFI}). It can be rewritten as
\begin{equation}
   \label{PolarFIright}
\mathcal{P}_{\sigma}(\tilde{\rho})\ \frac{2}{\tilde{\rho}}\int \limits _{  Diff_{+}^{1}([0,\,1]) }\,\delta \left(\frac{\varphi'(1)}{\varphi'(0)}-1\right)\ \varphi'(0)\ \mu_{\frac{2\sigma}{\tilde{\rho}}}(d\varphi)\,.
\end{equation}

After the substitution (\ref{subst}), the equation (\ref{PolarFIright}) takes the form
\begin{equation}
   \label{PolarXI}
\mathcal{P}_{\sigma}(\tilde{\rho})\ \frac{2}{\tilde{\rho}}\int\limits_{C_{0}([0,\,1])} \delta\left(\xi(1)\right)\,\frac{1}{\int \limits _{0}^{1}e^{\xi(\tau)}d\tau }\,w_{\frac{2\sigma}{\rho}}(d\xi)\,.
\end{equation}

To evaluate the integral (\ref{PolarXI}), we differentiate  the equation
$$
\int\limits_{C_{0}([0,\,1])} \delta\left(\xi(1)\right)\,\exp\left\{ \frac{-2\beta^{2}}{\kappa^{2}(\beta+1)}\frac{1}{\int \limits _{0}^{1}e^{\xi(\tau)}d\tau }\right\}\,w_{\kappa}(d\xi)
$$
\begin{equation}
   \label{XFormula}
  =\frac{1}{\sqrt{2\pi}\kappa}\,\exp \left\{-\frac{1}{2\kappa ^{2}}\left( 2\log (\beta +1)\,\right)^{2} \right\}
\end{equation}
proved in \cite{(BShCorrel)}.
Also we use the equality
$$
\delta \left(\tilde{\rho}-\left(\int \limits _{0}^{1}\,\frac{dt}{x^{2}(t)} \right)^{-\frac{1}{2}}\right)=\frac{2}{\tilde{\rho}^{3}}
\delta \left(\frac{1}{\tilde{\rho}^{2}}-\int \limits _{0}^{1}\,\frac{dt}{x^{2}(t)} \right)\,.
$$
 As a result,  (\ref{PolarFI}) is represented in the form
 \begin{equation}
   \label{P1}
\frac{1}{\sqrt{2\pi}\sigma} \mathcal{P}_{\sigma}(\tilde{\rho})=\frac{2}{\tilde{\rho}^{3}}\,\exp\left\{-\frac{\sigma^{2}}{8\tilde{\rho}^{2}}\right\}
 \int \limits _{ C_{+}([0,\,1]) }\delta \left(\frac{1}{\tilde{\rho}^{2}}-\int \limits _{0}^{1}\,\frac{dt}{x^{2}(t)} \right)
\delta\left(x(0)-x(1)\right)\,w_{\sigma}(dx)\,.
\end{equation}

To perform functional integration in the above equation, we take the Fourier transform of the first $\delta-$function in the integrand of (\ref{P1}), represent  the Wiener measure as
$$
w_{\sigma}(dx)=\exp\left\{-\frac{1}{2\sigma^{2}}\int \limits _{0}^{1}\left(x'(t) \right)^{2} dt\right\}\,dx\,,
$$
and substitute the variable $y(t)=\frac{x(t)}{\sigma}\,. $ As the result, the functional integral in the right-hand side of (\ref{P1}) has the form:
$$
\int \limits _{ C_{+}([0,\,1]) }\delta \left(\frac{1}{\tilde{\rho}^{2}}-\int \limits _{0}^{1}\,\frac{dt}{x^{2}(t)} \right)
\delta\left(x(0)-x(1)\right)\,w_{\sigma}(dx)
$$
\begin{equation}
   \label{Fourier}
=\frac{1}{2\pi}\int \limits _{-\infty}^{+\infty}\,d\tilde{\lambda}\,\exp\left\{i\frac{\tilde{\lambda}}{\tilde{\rho}^{2}}\right\}\int \limits _{  C_{+}([0,\,1]) }
\exp\left\{-\frac{1}{2}\int \limits _{0}^{1}\left[\left(y'(t) \right)^{2}+
\frac{2i\,\tilde{\lambda}}{\sigma^{2}y^{2}(t)}\right] dt\right\}\delta\left(y(0)-y(1)\right)dy\,.
\end{equation}
Re-scaling also the parameters $\tilde{\lambda}=\sigma^{2}\lambda\,,\ \tilde{\rho}^{2}=\sigma^{2}\eta^{2}\,, $ we rewrite (\ref{Fourier}) as
\begin{equation}
   \label{Fourier1}
\frac{\sigma^{2}}{2\pi}\int \limits _{-\infty}^{+\infty}\,d\lambda\,\exp\left\{i\frac{\lambda}{\eta^{2}}\right\}\int \limits _{  C_{+}([0,\,1]) }
\exp\left\{-\frac{1}{2}\int \limits _{0}^{1}\left[\left(y'(t) \right)^{2}+
\frac{2i\lambda}{y^{2}(t)}\right] dt\right\}\delta\left(y(0)-y(1)\right)\,dy\,.
\end{equation}

The integral over $ C_{+}([0,\,1]) $ in (\ref{Fourier1}) is nothing less than $Tr\,\exp\{-A_{0}\}\,,$ that is,
the functional integral for the partition function in the Calogero model given by the Euclidean action
\begin{equation}
   \label{ActCal}
A_{0}(g )=\frac{1}{2}\,\int \limits _{0}^{1}\left[\left(y'(t) \right)^{2}
+2g\frac{1}{y^{2}(t)}\right] dt\,,
\end{equation}
although with an imaginary coupling constant $g=i\lambda \,.$

Note that the appearance of the action of conformal quantum mechanics is tightly connected to the reduction of the Wiener measure to the orbit (\ref{inv}).

Since the action (\ref{ActCal}) has the continuous spectrum, we first consider the regularized action
\begin{equation}
   \label{ActOsc}
A_{\omega}(g )=\frac{1}{2}\,\int \limits _{0}^{1}\left[\left(y'(t) \right)^{2} +\omega^{2}\,y^{2}(t)
+2g\frac{1}{y^{2}(t)}\right] dt\,,
\end{equation}
calculate the Fourier transform
\begin{equation}
   \label{Fourier2}
=\frac{\sigma^{2}}{2\pi}\int \limits _{-\infty}^{+\infty}\,d\lambda\,\exp\left\{i\frac{\lambda}{\rho^{2}}\right\}\int \limits _{  C_{+}(S^{1}) }
\exp\left\{-A_{\omega}(g_{\lambda})\right\}\delta\left(y(0)-y(1)\right)\,dy\,,
\end{equation}
and then tend the parameter $\omega$ to zero.

The solution of the quantum problem for the action of the quantum oscillator with the square-inverse potential  (\ref{ActOsc}) is well known. The wave eigenfunctions $ \psi_{n}(x)\  (\psi_{n}(0)=0) $ form a basis in the Hilbert space of square integrable on the semiaxis $ 0<x<+\infty $ functions (see, e.g., \cite{(Perelomov)}) with the energy levels
$$
E_{n}=\frac{\omega}{2}\left(1+\sqrt{2g+\frac{1}{4}} \right)+2n\,\omega\,,\ \ \ \ \ \ \ n=0,\,1,\,2,\,...\,.
$$

Therefore the partition function for the action (\ref{ActOsc}) equals to
$$
\int \limits _{  C_{+}([0,\,1]) }\exp\{-A_{\omega}\}\delta\left(y(0)-y(1)\right)dy=Tr\exp\{-A_{\omega}\}
$$
\begin{equation}
   \label{PF}
=\sum \limits _{0}^{\infty}\exp\left\{-E_{n} \right\}=\exp\left\{\frac{\omega}{2}\left[1-\sqrt{2g+\frac{1}{4}}\right] \right\}\frac{1}{2\sinh \omega}\,.
\end{equation}

The integral over $ \lambda$ in (\ref{Fourier1}) has the form
\begin{equation}
   \label{IntLAMB}
\int \limits _{-\infty}^{+\infty}\,d\lambda\,\exp\left\{i\frac{\lambda}{\eta^{2}}\right\}\exp\left\{-\frac{\omega}{2}\sqrt{2i\lambda+\frac{1}{4}} \right\}
=\sqrt{2\pi}\,\omega\,\eta^{3}\,\exp\left\{-\frac{1}{8\eta^{2}}-\frac{\omega^{2}}{8}\eta^{2} \right\}\,.
\end{equation}
To find the integral (\ref{IntLAMB}), we first substitute $\zeta=2i\lambda+\frac{1}{4}\,, $ divide the integral
$$
\int\limits_{Re \zeta=0}=
\int\limits_{Re \zeta=0,\,Im \zeta\geq 0}+\int\limits_{Re \zeta=0,\,Im \zeta<0}
=\int\limits_{Re \zeta\leq 0,\,Im \zeta= 0}+\int\limits_{Re \zeta>0,\,Im \zeta=0}\,,
$$
and make the substitutions $\zeta=i\sqrt{z}\,,\ \zeta=-i\sqrt{z} $ in the first and in the second integrals correspondingly.

After taking the limit $\omega\rightarrow 0$ in the above equations, from (\ref{P1}) we obtain
\begin{equation}
   \label{P2}
 \mathcal{P}_{\sigma}(\tilde{\rho})=\exp\left\{-\frac{\sigma^{2}}{8\tilde{\rho}^{2}}\right\}
 \,.
\end{equation}

Thus we prove the polar decomposition of the Wiener measure
\begin{equation}
   \label{PolarInterval}
w_{\sigma}(dx)=\exp\left\{-\frac{\sigma^{2}}{8\rho^{2}}\right\}\ \left(\varphi'(0)\varphi'(1) \right)^{\frac{3}{4}}\  \mu_{\frac{2\sigma}{\rho}}(d\varphi)\ d\rho\,,
\end{equation}
$$
x\in C_{+}([0,\,1])
\,,
\ \ \ \ \ \varphi\in Diff^{1}_{+}([0,\,1])\,,
\ \ \ \ \ 0<\rho<+\infty\,.
$$

The polar decomposition is  also valid for the Wiener measure on the space $C_{+}(S^{1})$. However in this case,
to evaluate the factor $\mathcal{P}(\rho)$ we are to normalize the measure in some way. For example, we can parametrize the circle of unit length $S^{1}$
by the interval $[0,\,1]\,, $ and consider the measure on  $C_{+}(S^{1})$ as the measure on  $C_{+}([0,\,1])$ with the ends of the interval "glued":
$ x(0)=x(1)\,.$ In this case, it has the form
\begin{equation}
   \label{PolarCircle}
w_{\sigma}(dx)=\exp\left\{-\frac{\sigma^{2}}{8\rho^{2}}\right\}\ \mu_{\frac{2\sigma}{\rho}}(d\varphi)\ d\rho\,,
\end{equation}
$$
x\in C_{+}(S^{1})
\,,
\ \ \ \ \ \varphi\in Diff^{1}_{+}(S^{1})\,,
\ \ \ \ \ 0<\rho<+\infty\,.
$$

\section{Connection between functional integrals in the Schwarzian theory and in conformal quantum mechanics }
\label{sec:conSchw}

From the polar decomposition of the Wiener measure (\ref{PolarCircle} ), we have the equality for the functional integrals
$$
\int \limits _{  C_{+}(S^{1}) }F(x)\,
\exp\left\{-\frac{1}{2\sigma^{2}}\int \limits _{S^{1}}\left[\left(x'(t) \right)^{2} -2\pi^{2}x^{2}(t) +
\frac{2g}{x^{2}(t)}\right] dt\right\}\,dx
$$
$$
=\int\limits_{0}^{+\infty}\,d\rho\,\exp\left\{-\left(\frac{\sigma^{2}}{8}+\frac{g}{\sigma^{2}}\right)\frac{1}{\rho^{2}} \right\}
$$
\begin{equation}
   \label{FIeq}
\times\int \limits _{  Diff_{+}^{1}(S^{1}) }F(x(\rho,\,\varphi))\,
\exp\left\{\frac{\rho^{2}}{4\sigma^{2}}\int \limits _{S^{1}}\left[\mathcal{S}_{\varphi}(t)+2\pi^{2}\left(\varphi'(t)\right)^{2}
\right] dt\right\}\,d\varphi\,.
\end{equation}

In particular, for
$$
F(x)=\Phi(x)\,\exp\left\{-\frac{\beta^{2}}{\sigma^{2}}\int \limits _{S^{1}}x^{2}(t)dt
 \right\}
$$
with $\Phi(x)$ good enough, the integrals are well defined. Therefore we have a transparent connection between conformal quantum mechanics and the Schwarzian theory.

If the functional $\Phi$ is  $SL(2,\textbf{R}) $ invariant then it can be reduced to the orbits. In this case, the integrals can be factorized
with the result
$$
\int \limits _{  C_{+}(S^{1})/SL(2,\textbf{R}) }\Phi(x)\,
\exp\left\{-\frac{1}{2\sigma^{2}}\int \limits _{S^{1}}\left[\left(x'(t) \right)^{2} -2\pi^{2}x^{2}(t) +
\frac{2g}{x^{2}(t)}\right] dt\right\}\,dx
$$
$$
=\int\limits_{0}^{+\infty}\,d\rho\,\exp\left\{-\left(\frac{\sigma^{2}}{8}+\frac{g}{\sigma^{2}}\right)\frac{1}{\rho^{2}} \right\}
$$
\begin{equation}
   \label{FIfac}
\times\int \limits _{  Diff_{+}^{1}(S^{1})/SL(2,\textbf{R}) }\Phi(x(\rho,\,\varphi))\,
\exp\left\{\frac{\rho^{2}}{4\sigma^{2}}\int \limits _{S^{1}}\left[\mathcal{S}_{\varphi}(t)+2\pi^{2}\left(\varphi'(t)\right)^{2}
\right] dt\right\}\,d\varphi\,.
\end{equation}

Evaluating functional integral in conformal quantum mechanics, we thus find the corresponding functional integral in the Schwarzian theory, and vice versa,
the technique of functional integration over the group of diffeomorphisms makes it possible to evaluate nontrivial functional integrals in conformal quantum mechanics.

Consider a simple example. Let us put $g=0,\ \sigma=1 $ and take
$$
\Phi(x)= \delta \left(\kappa^{2}-4\int \limits _{S^{1}}\,\frac{dt}{x^{2}(t)} \right)\,.
$$
The right-hand side defines the partition function in the Schwarzian theory
$$
 Z_{Schw}(\kappa)=\frac{1}{\sqrt{2\pi}\kappa}\int \limits _{ Diff_{+}^{1}(S^{1})/SL(2,\textbf{R}) }\exp\left\{-I \right\}  d\varphi\,.
$$
The left-hand side of (\ref{FIfac}) leads  to the result \cite{(BShUnusual)}
$$
 Z_{Schw}(\kappa)=\frac{1}{\sqrt{2\pi}\kappa^{3}}\,\exp\left\{\frac{2\pi^{2}}{\kappa^{2}} \right\}
$$
obtained early \cite{(BShExact)} by direct functional integration in the Schwarzian theory.

In the same way, one can use functional integrals in conformal quantum mechanics to evaluate other Schwarzian functional integrals.

\section{An example of application of the polar decomposition in conformal quantum mechanics}
\label{sec:Anex}

In \cite{(BShUnusual)}, we used the exact solutions of the quantum oscillator model with the Calogero potential to evaluate the functional integral
 assigning the partition function in the Schwarzian theory.

The polar decomposition of the Wiener measure (\ref{PolarInterval}) can be also used in the opposite direction. Namely, we can use already evaluated functional integrals of the Schwarzian theory to perform nontrivial functional integration in conformal quantum mechanics.
In this way, we find the fundamental solution of the Schr\"{o}dinger equation in imaginary time in the quantum mechanical model with Calogero-type potential.

Consider now the heat transfer equation of the form
\begin{equation}
   \label{HTeq}
\frac{\partial}{\partial\tilde{\tau}}\psi_{g}(q,\,\tilde{\tau})=\left(\frac{1}{2}\frac{\partial^{2}}{\partial q^{2}}-\frac{g}{q^{2}}\right)\,\psi_{g}(q,\,\tilde{\tau})\,,\ \ \ \ \ q>0\,.
\end{equation}
It is the Schr\"{o}dinger equation in imaginary time with the potential $V(q)=\frac{g}{q^{2}}\,. $

To exclude the paths with negative values of $x(\tau)$ in the subsequent functional integrals, we put the boundary condition
$
\psi_{g}(q=0,\,\tilde{\tau})=0\,,
$
and integrate over the space $ C_{+}([0,\,t])\,.$

Let the initial condition be
$
\psi_{g}(q,\,0)=\psi_{0}(q)=\delta (q(0)-q_{0})\,,\ \ \ \ \ q_{0}>0\,.
$

The fundamental solution of the Cauchy problem is given by the functional integral
 \begin{equation}
   \label{FundSol}
\psi_{g}(t,\,q,\, q_{0})=\int \limits _{ C_{+}([0,\,t]) }\delta(x(0)-q_{0})\,\delta(x(t)-q)\,\exp
\left\{-\int \limits _{0}^{t}\,\frac{g}{x^{2}(\tilde{\tau})} \,d\tilde{\tau}\right\}\,w_{1}(dx)
\,.
\end{equation}
After the substitution
$$
\tilde{\tau}=t\,\tau\,,\ \ \ y(\tau)=x(t\,\tau)\,,\ \ \ \tau\in [0,\,1]\,,
$$
it turns to
\begin{equation}
   \label{FundSoly}
\psi_{g}(q,\,t\,;\ q_{0})=\int \limits _{ C_{+}([0,\,1]) }\delta(\sqrt{t}\,y(0)-q_{0})\,\delta(\sqrt{t}\,y(1)-q)\,\exp
\left\{-\int \limits _{0}^{1}\,\frac{g}{y^{2}(\tau)} \,d\tau\right\}\,\sqrt{t}\,w_{1}(dy)
\,.
\end{equation}

Now we use the polar decomposition of the Wiener measure (\ref{PolarInterval}) and get
$$
\psi_{g}(q,\,t\,;\ q_{0})=\frac{1}{\sqrt{t}}\int \limits _{0}^{+\infty}\,d\rho,\exp\left\{-\frac{1}{\rho^{2}}\left(g+\frac{1}{8} \right) \right\}
$$
\begin{equation}
   \label{FundSolmu}
\times \int \limits _{ Diff^{1}_{+}([0,\,1]) }\delta\left(\rho\sqrt{\varphi'(0)}-\frac{q_{0}}{\sqrt{t}}\right)\,\delta\left(\rho\sqrt{\varphi'(1)}-\frac{q}{\sqrt{t}}\right)\ \left(\varphi'(0)\varphi'(1) \right)^{\frac{3}{4}}\ \mu_{\frac{2}{\rho }}(d\varphi)
\end{equation}

In terms of the special functional integral
\begin{equation}
   \label{E}
\mathcal {E}_{\sigma}(u,\,v)=\int\limits_{Diff^{1}_{+} ([0,1]) }\,\delta\left(\varphi'(0)-u \right)\,\delta\left(\varphi'(1)-v \right)\,\mu_{\sigma}(d\varphi)\,,
\end{equation}
it is written as
$$
\psi_{g}(q,\,t\,;\ q_{0})=\frac{1}{\sqrt{t}}\int \limits _{0}^{+\infty}\,d\rho\,\exp\left\{-\frac{1}{\rho^{2}}\left(g+\frac{1}{8} \right) \right\}
$$
$$
 \times  \int \limits _{0}^{+\infty}\,du\,\int \limits _{0}^{+\infty}\,dv\,\delta\left(\rho\sqrt{u}-\frac{q_{0}}{\sqrt{t}}\right)\,\delta\left(\rho\sqrt{v}-\frac{q}{\sqrt{t}}\right)\ \left(u\,v \right)^{\frac{3}{4}}\ \mathcal {E}_{\frac{2}{\rho }}(u,\,v)
$$
\begin{equation}
   \label{FundSolE}
=\frac{4}{t^{3}}\left(q_{0}q\right)^{\frac{5}{2}}\int \limits _{0}^{+\infty}\,d\rho\,\frac{1}{\rho^{7}}\,\exp\left\{-\frac{1}{\rho^{2}}\left(g+\frac{1}{8} \right) \right\}\,
\mathcal {E}_{\frac{2}{\rho } }\left(\frac{q_{0}^{2}}{t\,\rho^{2}},\,\frac{q^{2}}{t\,\rho^{2}}\right)
\,.
\end{equation}

In \cite{(BShCorrel)}, we performed the functional integration in (\ref{E})  and represented
the function $\mathcal {E}_{\sigma }\left(u,\,v\right) $ in the form of the ordinary integral:
$$
\mathcal {E}_{\sigma }\left(u,\,v\right)
=\left(\frac{2}{\pi\sigma^{2}}\right)^{\frac{3}{2}}\,\frac{1}{\sqrt{uv}}\,\exp\left\{\frac{2}{\sigma^{2}}\left(-u-v\right) \right\}
$$
\begin{equation}
   \label{Eint}
\times\int\limits_{0}^{+\infty}\,\exp\left\{-\frac{2}{\sigma^{2}}\left(2\,\sqrt{u\,v}\,\cosh\theta+\theta^{2}-\pi^{2}\right) \right\}\,\sin\left(\frac{4\pi\theta}{\sigma^{2}} \right)\,\sinh\theta\,d\theta \,.
\end{equation}

With the explicit form of $\mathcal {E}_{\sigma}(u,\,v) $ given by (\ref{Eint}), the equation (\ref{FundSolE}) is transformed into
$$
\psi_{g}(q,\,t\,;\ q_{0})=\frac{1}{4i}\,\frac{4}{t^{2}}\left(\frac{q_{0}\,q}{2\pi }\right)^{\frac{3}{2}}\exp\left\{-\frac{q_{0}^{2}+q^{2}}{2t} \right\} \int \limits _{0}^{+\infty}\,d\rho\,\frac{1}{\rho^{2}}\,\exp\left\{-\frac{1}{\rho^{2}}\left(g+\frac{1}{8} \right) \right\}\,
$$
 \begin{equation}
   \label{FundSoltheta}
\times\int \limits _{-\infty}^{+\infty}d\theta\,\sinh \theta\,\exp
\left\{-\frac{q_{0}\,q}{t}\cosh\theta\right\}
\left[\exp
\left\{-\frac{\rho^{2}}{2}\left(\theta-i\pi\right)^{2} \right\}- \exp
\left\{-\frac{\rho^{2}}{2}\left(\theta+i\pi\right)^{2} \right\} \right]
\,.
\end{equation}

Substituting $z=\theta-i\pi$ or $z=\theta+i\pi$,
we rewrite the integral  over $\theta$ as
$$
 \lim\limits_{R\rightarrow +\infty}\left[\int \limits _{-R+i\pi}^{+R +i\pi}\,f(z)\,dz-\int \limits _{-R+i\pi}^{+R +i\pi}\,f(z)\,dz\right]
$$
 \begin{equation}
   \label{f}
=\lim\limits_{R\rightarrow +\infty}\left[\int \limits _{R-i\pi}^{R +i\pi}\,f(z)\,dz-\int \limits _{-R-i\pi}^{-R +i\pi}\,f(z)\,dz\right]
=2i\lim\limits_{R\rightarrow +\infty}\,\int \limits _{-\pi}^{+\pi}\,f(R+i\tau)\,d\tau\,,
\end{equation}
where
$$
f(z)=\sinh z\,\exp
\left\{\frac{q_{0}\,q}{t}\cosh z\right\}
\exp
\left\{-\frac{\rho^{2}}{2}z^{2} \right\}\,.
$$

Thus we have
$$
\psi_{g}(q,\,t\,;\ q_{0})=\frac{2}{t^{2}}\left(\frac{q_{0}\,q}{2\pi }\right)^{\frac{3}{2}}\exp\left\{-\frac{q_{0}^{2}+q^{2}}{2t} \right\} \int \limits _{0}^{+\infty}\,d\rho\,\frac{1}{\rho^{2}}\,\exp\left\{-\frac{1}{\rho^{2}}\left(g+\frac{1}{8} \right) \right\}\,
$$
 \begin{equation}
   \label{psiLim}
\times\lim\limits_{R\rightarrow +\infty}\,\int \limits _{-\pi}^{+\pi}
\,\exp
\left\{\frac{q_{0}\,q}{t}\cosh(R+i\tau)\right\}
\exp
\left\{-\frac{\rho^{2}}{2}\left(R+i\tau\right)^{2} \right\}\,
\sinh (R+i\tau)\,d\tau\,.
\end{equation}

Now we change the order of the integrals and integrate over $\rho\,.$ Then after the substitution $\zeta=\cosh(R+i\tau)\,,$ we take the limit
$R\rightarrow +\infty\,.$ As the result, (\ref{psiLim}) is transformed into
 \begin{equation}
   \label{zeta}
\psi_{g}(q,\,t\,;\ q_{0})=\frac{(q_{0}\,q)^{\frac{3}{2}}}{\pi t^{2}}\frac{1}{\sqrt{2g+\frac{1}{4}}}\exp\left\{-\frac{q_{0}^{2}+q^{2}}{2t} \right\}\,\frac{1}{2i}
\left[\int \limits _{\Gamma_{2}}
\chi_{g}(\zeta)\,d\zeta -
\int \limits _{\Gamma_{3}}\chi_{g}(\zeta)\,
d\zeta
\right],
\end{equation}
where
$$
\chi_{g}(\zeta)=\exp
\left\{\frac{q_{0}\,q}{t}\,\zeta\right\}
\exp
\left\{- \sqrt{2g+\frac{1}{4}}\,arc\cosh\zeta\right\}
\,,
$$
$\Gamma_{2}$ and $\Gamma_{3}$ are the lower and the upper edges of the cut along the real axis in the complex plane of $\zeta$ from $-\infty$ to $1\,.$

Note that for $-1<x<1 $
$$
arc\cosh(x-i0)=-i\arccos x\,,\ \ \ \ \ arc\cosh(x+i0)=i\arccos x\,,
$$
and for $-\infty<x<-1 $
$$
arc\cosh(x-i0)=\log(|x|+\sqrt{x^{2}-1})-i\pi\,,\ \ \  \ arc\cosh(x+i0)=\log(|x|+\sqrt{x^{2}-1})+i\pi\,.
$$

After corresponding reorganization of the integrals in (\ref{zeta}) and obvious substitutions,
we obtain  the fundamental solution of the Cauchy problem in the final form
$$
 \psi_{g}(q,\,t\,;\ q_{0})=\mathcal {F}_{g}(q,\,t\,;\,q_{0})\,\sin\left( \pi\,\sqrt{2g+\frac{1}{4}}\right)
$$
$$
 \times\,\int \limits _{0}^{+\infty}\,\exp
\left\{-\frac{q_{0}\,q}{t} \,\cosh\theta \right\}\,\exp\left\{-\sqrt{2g+\frac{1}{4}}\, \theta \right\}\,\sinh \theta\ d\theta
 $$
 \begin{equation}
   \label{FundSolFinal}
+\ \mathcal {F}_{g}(q,\,t\,;\,q_{0})\, \int \limits _{0}^{\pi}\,\exp
\left\{\frac{q_{0}\,q}{t}\, \cos\tau \right\}\,\sin\left( \sqrt{2g+\frac{1}{4}}\,\tau\right)\,\sin\tau\,d\tau
\,,
\end{equation}
where
\begin{equation}
   \label{Fmath}
\mathcal {F}_{g}(q,\,t\,;\,q_{0})\equiv\frac{(q_{0}\,q)^{\frac{3}{2}}}{\pi\, t^{2}}\,\frac{1}{\sqrt{2g+\frac{1}{4}}}\,\exp\left\{-\frac{q_{0}^{2}+q^{2}}{2t} \right\}\,.
\end{equation}

If $g=0$, the integrals can be evaluated explicitly. In this case,
$$
\chi_{0}(\zeta)=\exp
\left\{\frac{q_{0}\,q}{t}\,\zeta\right\}
\exp
\left\{- \frac{1}{2}\,arc\cosh\zeta\right\}
\,.
$$
Representing
$$
\exp
\left\{- \frac{1}{2}\,arc\cosh\zeta\right\}=\cosh\left\{- \frac{1}{2}\,arc\cosh\zeta\right\}-\sinh\left\{- \frac{1}{2}\,arc\cosh\zeta\right\}=\sqrt{\frac{\zeta+1}{2}}-\sqrt{\frac{\zeta-1}{2}}\,,
$$
and making substitutions
$$
\xi=i\sqrt{\frac{\zeta+1}{2}}\,,\ \eta=i\sqrt{\frac{\zeta-1}{2}}
$$
in the integrals (\ref{zeta}), we get the expected result
\begin{equation}
   \label{g0res}
\psi_{0}(q,\,t\,;\ q_{0})=\frac{1}{\sqrt{2\pi t}}\,\left[ \exp\left\{-\frac{(q_{0}-q)^{2}}{2t} \right\}-\exp\left\{-\frac{(q_{0}+q)^{2}}{2t} \right\}\right]\,.
\end{equation}

\section{Concluding remarks}
\label{sec:concl}

 In this paper, we demonstrate that,
due to the existence of the $Diff_{+}^{3}$-invariant (\ref{inv}), there is the one-to-one correspondence
$$
C_{+}([0,\,1])\leftrightarrow (0,\,+\infty)\times Diff^{1}_{+} ([0,\,1])\,,
$$
and the Wiener measure  can be written as (\ref{PolarInterval}).
The factor (\ref{P2}) determines the relative weight of the paths $x(t) $ with a definite value of the invariant (\ref{inv}).

Having in mind the analogy with the form of the Riemann-Lebesgue measure on two-dimensional plane in polar co-ordinates, we give the name "polar decomposition of the Wiener measure" to the  representation (\ref{PolarInterval}).
Elements of the group $\varphi\in Diff^{1}_{+}$ play the role of angles, and values of the invariant $\rho$ correspond to lengths of radius vectors.

The equation (\ref{FIfac}) connecting Schwarzian functional integrals with those in conformal quantum
mechanics
gives us the possibility to choose the most successful strategy of functional integration.

After the substitution (\ref{subst}) we can rewrite the polar decomposition of the Wiener measure (\ref{PolarInterval} ) as
\begin{equation}
   \label{PolarW}
w_{\sigma}(dx)=\exp\left\{-\frac{\sigma^{2}}{8\rho^{2}}\right\}\ \frac{e^{\frac{3}{4}\xi(1)}}{\left(\int \limits _{0}^{1}\,e^{\xi(\tau)}d\tau\right) ^{\frac{3}{2}}}\  w_{\frac{2\sigma}{\rho}}(d\xi)\ d\rho\,.
\end{equation}

In \cite{(BShRules)}, we have already noticed the apparent violation of  Markov property by the function $\varphi(t)\,. $
Although $x(t)$ and $\xi (\tau)$ are both Wiener processes , the Markov behaviour of $x(t)$ with respect to the time $t$ of "its own world"  obviously does not imply  its Markov behaviour with respect to the time $\tau $ of the "shadow world", and vice versa.

\end{document}